\documentclass[12pt,a4paper,twoside]{article}
\usepackage{graphics}
\usepackage{color}

\usepackage[dvips]{epsfig}
\usepackage{amstext,amsmath,amssymb}

\newcommand{\half}{{\textstyle\frac{1}{2}}}
\def\Re{\mathop{\mathrm{Re}}\nolimits}
\def\Im{\mathop{\mathrm{Im}}\nolimits}

\newlength{\figwidth}
\newlength{\figheight}
\def\z0{Z}

\title{
       \vspace*{-1.5cm}
       \begin{flushright}
       \begin{tabular}{l}
       {\normalsize CERN-TH/2002-330 }    \\[-3mm]
       {\normalsize IFT - 40/2002}    \\[-3mm]
       {\normalsize hep-ph/0211371}\\
       {\normalsize November 2002}
       \end{tabular}
       \end{flushright}
       \vspace*{1.5cm}
      \sc  
Two-Higgs-Doublet Models \protect{\\}
with CP violation
\thanks{Presented at SUSY02, DESY, June 2002 and 
LCWS02, Jeju, Korea, August 2002}}

\author{I.~F.\ Ginzburg$^1$,
        M.\ Krawczyk$^{2,3}$ \ and 
        P.\ Osland$^4$
\\
\\
        $^1$ {\it Sobolev Institute of Mathematics, SB RAS,
        630090 Novosibirsk, Russia}
\\
       $^2$ {\it Theory Division CERN, CH-1211 Geneva 23, Switzerland}
\\
        $^3$ {\it Institute of Theoretical Physics, Warsaw University, Poland}
\\
        $^4$ {\it Department of Physics, University of Bergen,
        N-5007 Bergen, Norway}
}

\date{}

\begin{document}
		
\maketitle

\vfill

\begin{abstract}
We consider the Two-Higgs-Doublet Model and determine the range of
parameters for which CP violation and Flavor Changing Neutral Current effects
are naturally small.  It corresponds to small values of the mass parameter
$m_{12}^2$, describing soft ($\phi_1,\,\phi_2$) mixing in the potential.  We
discuss how, in this approach, some Higgs bosons can be heavy, with mass of
the order of 1~TeV.

The possibility that at the Tevatron, LHC and an $e^+e^-$ Linear Collider,
only one Higgs boson will be found, with properties indistinguishable from
those in the Standard Model (SM), we define as the SM-like scenario.  While
this scenario can be obtained with large $\mu^2 \sim \Re m_{12}^2$ parameter,
in which case there is decoupling, we here discuss the opposite case of small
$\mu^2$, without decoupling.
\end{abstract}

\vfill
\begin{flushleft}
CERN-TH/2002-330 \\
IFT - 40/2002    \\
hep-ph/0211371 \\
November 2002
\end{flushleft}
\thispagestyle{empty}
\clearpage

\section{Introduction}

We consider the following Two-Higgs-Doublet Model (2HDM) potential, with
quartic and quad\-ratic terms separated
\cite{Diaz-Cruz:1992uw,Haber:1994mt,Branco,Ginzburg:2001ss,Gunion:2002zf}:
\begin{align}  \label{Eq:pot}
V&=\half\lambda_1(\phi_1^\dagger\phi_1)^2
+\half\lambda_2(\phi_2^\dagger\phi_2)^2 
+\lambda_3(\phi_1^\dagger\phi_1) (\phi_2^\dagger\phi_2)
+\lambda_4(\phi_1^\dagger\phi_2) (\phi_2^\dagger\phi_1) \nonumber \\
&+\half\bigl[{\lambda_5}(\phi_1^\dagger\phi_2)^2+{\rm h.c.}\bigr]
+\bigl\{\bigl[\lambda_6(\phi_1^\dagger\phi_1)
+\lambda_7(\phi_2^\dagger\phi_2)\bigr](\phi_1^\dagger\phi_2)
+{\rm h.c.}\bigr\} \nonumber\\
&-\bigl\{m_{11}^2(\phi_1^\dagger\phi_1)
+{\bigl[m_{12}^2 (\phi_1^\dagger\phi_2)+{\rm h.c.}\bigr]}
+m_{22}^2(\phi_2^\dagger\phi_2)\bigr\}.
\end{align}

As is well known, both CP violation in the Higgs sector and
flavor-changing neutral currents (FCNC) can be
suppressed by imposing a $Z_2$ symmetry \cite{Glashow:1976nt}.
This requires symmetry of the potential under 
($\phi_1\to-\phi_1,\; \phi_2\to\phi_2$) (or vice versa),
which implies $\lambda_6=\lambda_7=m_{12}^2=0$.
We shall allow {\it soft}\ violation of this symmetry, i.e., we take
$\lambda_6=\lambda_7=0$, but allow $m_{12}^2\ne 0$ \cite{Branco,
Gunion:2002zf,Santos:2001tp}.
A simple discussion can be given for this case, in which
$\Im m_{12}^2 \neq 0$ signals CP violation.

\section{\boldmath$CP$ violation}

We shall now consider the simpler case of $\lambda_6=\lambda_7=0$, 
and parametrize the minimum of the potential (or vacuum) as
\begin{equation}
\phi_1=\begin{bmatrix}
0 \\
\frac{1}{\sqrt{2}}\,v_1
\end{bmatrix},\quad
\phi_2=\begin{bmatrix}
0 \\
\frac{1}{\sqrt{2}}\,v_2\,e^{i\xi}
\end{bmatrix}.
\end{equation}
Naively, the phase $\xi$ violates $CP$, but it can be removed by a global
phase transformation on the field $\phi_2$, together with the phases
of $\lambda_5$, $m_{12}^2$ and the fermion fields \cite{Branco,GKO}.
It is convenient to define
\begin{equation}  \label{Eq:musq}
\mu^2={\Re (m_{12}^2 e^{i\xi})}\frac{v^2} {v_1v_2}.
\end{equation}
The phase $\xi$ can be found from the equation
\begin{equation}
\Im(m_{12}^2\,e^{i\xi})
=\Im(\lambda_5\,e^{2i\xi})v_1v_2. \label{constr}
\end{equation}
Making use of the rephasing invariance \cite{Branco,GKO}, 
we put $\xi=0$. With this choice, eq.~(\ref{constr}) becomes
a constraint for the relation of $\Im(m_{12}^2)$ to $\Im\lambda_5$.

The neutral sector has a mass squared matrix of the form
\begin{equation} \label{Eq:mass-squared}
{\cal M}^2=\left(\begin{array}{ccc}
{\cal M}_{11}^2&
{\cal M}_{12}^2&
-\half\Im\lambda_5v^2\sin\beta\\[4pt]
{\cal M}_{12}^2&
{\cal M}_{22}^2&
-\half\Im\lambda_5v^2\cos\beta\\[4pt]
-\half\Im\lambda_5v^2\sin\beta& 
-\half\Im\lambda_5v^2\cos\beta&
{\cal M}_{33}^2
\end{array}\right)
\end{equation}
where ${\cal M}_{11}^2$, ${\cal M}_{12}^2$, ${\cal M}_{22}^2$ and
${\cal M}_{33}^2$ are the same as
in the CP-conserving case. When $\Im\lambda_5=0$, there is no $CP$
violation, the matrix (\ref{Eq:mass-squared}) is block diagonal, and the
physical states are $h$, $H$ and $A$. When $\Im\lambda_5\ne 0$, all three
neutral Higgs states mix; we denote them by $h_1$, $h_2$ and $h_3$.  

The mass-squared matrix may be diagonalized via a rotation matrix,
defined by
\begin{equation} \label{Eq:R}
R\,{\cal M}^2\,R^{\rm T}=\text{diag}(M_1^2,M_2^2,M_3^2).
\end{equation}
In the limit of weak $CP$ violation, the masses for $h_1$, $h_2$ and $h_3$
will deviate from those of $h$, $H$ and $A$ by 
terms quadratic in $\Im\lambda_5$.

\section{Decoupling or no decoupling?}

We shall here consider the scenario of weak (or no) $CP$ violation
and large masses of Higgs particles except one, namely $M_{h_1} \sim M_h$. 
Let us discuss how
large masses $M_{A}$ (close to $M_{h_3}$) 
and  $M_{H^{\pm}}$ arise in such a case. 
The potential (\ref{Eq:pot})
(but with $\lambda_6=\lambda_7=0$) gives
\begin{equation}
M_A^2=\half\mu^2-\Re\lambda_5v^2\quad \text{and}\quad
M_{H^{\pm}}^2=\half[\mu^2-(\lambda_4+\Re\lambda_5)v^2].
\end{equation}
There are two rather distinct mechanisms for obtaining large mass 
 $M_A^2$ and 
 $M_{H^{\pm}}^2$: either {\it (i)} $\mu^2$ is large (this is extensively
discussed by Haber as {\it the decoupling scenario})
\cite{Haber:1994mt,Gunion:2002zf}, or {\it (ii)} $\mu^2$ is small, whereas
$|\Re\lambda_5|$ is ``large'' \cite{Ginzburg:2001ss,Gunion:2002zf}.  In the
latter case, there are obvious upper bounds (from perturbativity and
positivity) on how large $|\Re\lambda_5|$ can be.
Decoupling properties of the Two-Higgs-Doublet Model were studied in 
\cite{Ciafaloni:1996ur}.

In this model, with weak (or no) $CP$ violation,
one can realize a Standard-Model-Like Scenario:
\begin{itemize}
\itemsep=-2pt
\it
\item
There is a light Higgs boson with couplings to the up (e.g.~$t$) and down 
(e.g.~$b$) type 
quarks, and to $W$ and $Z$, like in the Standard Model, 
\begin{equation}
|g_i|\approx|g_i^{\rm SM}| \quad (i=W,Z, \text{down}, \text{up}).
\end{equation}
\item
The other Higgs bosons are heavy, ${\cal O}(1~\text{TeV})$.
\end{itemize}

Within the Two-Higgs-Doublet Model, this scenario can be realized in
two distinct ways. They are \cite{Ginzburg:2001ss}:
\begin{itemize}
\itemsep=-2pt
\item[$-$]
Solutions~A. All basic couplings are approximately the same as in the SM,
up to an overall sign.
\item[$-$]
Solutions~B. Like Solutions~A, {\it except that
the couplings to either up- or down-type quarks have opposite signs
of those in the SM.} This case cannot be realized in
the decoupling scenario.
\end{itemize}
\section{Model II. Observables}
Let us now be more specific, and consider the so-called Model~II for Yukawa
couplings, where masses of down- and up-type quarks originate from couplings
to $\phi_1$ and $\phi_2$, respectively. 
We denote by $\chi_V^{h_i}$,
$\chi_u^{h_i}$ and $\chi_d^{h_i}$ the ratios of the Higgs
couplings to $W$ and $Z$ ($V$) and to up and down-type
quarks, with respect to those of the Standard Model.
In particular, for the Yukawa couplings these ratios can be
expressed via elements of the rotational matrix $R$ of
eq.~(\ref{Eq:R}) as
\begin{equation}
\chi_u^{h_i}=\frac{1}{\sin\beta}[R_{i2}-i\gamma_5\cos\beta R_{i3}],
\qquad
\chi_d^{h_i}=\frac{1}{\cos\beta}[R_{i1}-i\gamma_5\sin\beta R_{i3}],
\label{fermcoupl}
\end{equation}
where $R_{i3}$ is proportional to $\Im\lambda_5$.
Note that in accordance with eq.~(\ref{fermcoupl}), the $CP$
violation induced by Higgs exchange in $t\bar t$ production
\cite{Bernreuther:1993hq} provides information on $\Im\lambda_5$.

Furthermore, these relative couplings satisfy a {\it pattern relation}
\cite{Ginzburg:2001ss,GKO}:
\begin{equation}
(\chi_u^{h_i}+\chi_d^{h_i})\chi_V^{h_i}=1+\chi_u^{h_i}\chi_d^{h_i}.
\end{equation}

In the CP-conserving case,
even with all basic couplings being the same (up to a sign) as in the SM (8),
loop-induced transition rates, like $h\to\gamma\gamma$,
may differ from the SM prediction.
This is due to the different behaviors
of the trilinear Higgs coupling $hH^+H^-$ for small and large $\mu$.
In fact, the ratio of this coupling to its SM value can be written as
\begin{equation}
\chi_{H^\pm}^h \equiv-\frac{vg_{h H^+H^-}}{2M_{H^\pm}^2}
=\left(1-\frac{M_h^2}{2M_{H^\pm}^2}\right)\chi_V^{h}
+\frac{M_h^2-\half\mu^2}{2M_{H^\pm}^2}
(\chi_u^{h}+\chi_d^{h}).
\end{equation}
Thus, if $\mu^2\sim M_{H_\pm}^2$, there is no effect in
$\Gamma_{\gamma\gamma}$, whereas if $\mu^2< M_{H_\pm}^2$ there is a difference
of several per cent, as illustrated in Fig.~1 (left panel) for the case of 
Solutions~A \cite{Ginzburg:2001ss}.
Non-decoupling effects in the 2HDM were studied for
other processes in \cite{Malinsky:2002mq,Kanemura:2002cc}.
\begin{figure}
\refstepcounter{figure}
\label{Fig:lvls-05}
\addtocounter{figure}{-1}
\begin{center}
\setlength{\unitlength}{1cm}
\begin{picture}(15.5,6.)
\put(1.5,0.0){
\mbox{\epsfysize=6.0cm\epsffile{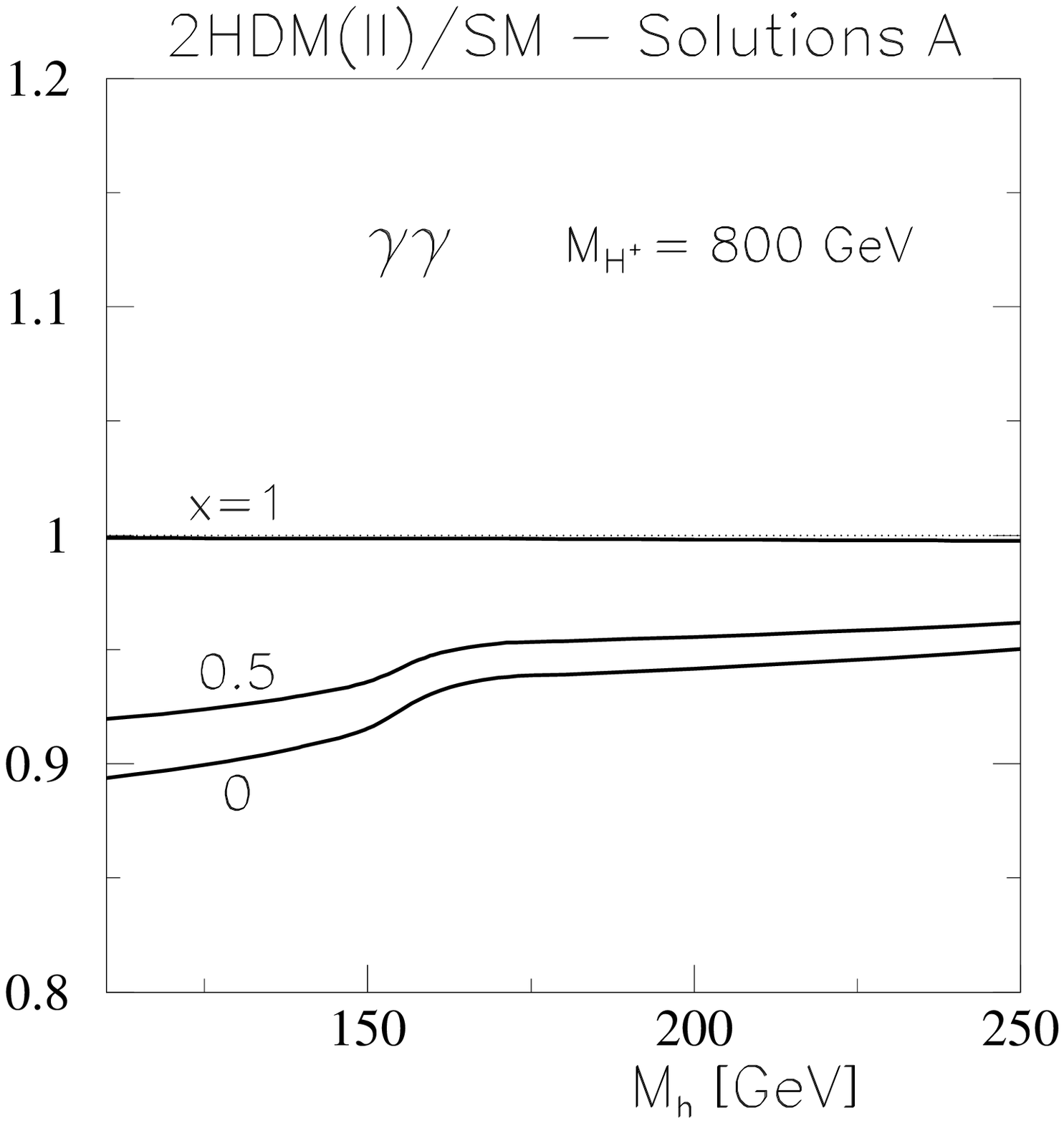}}
\mbox{\epsfysize=6.0cm\epsffile{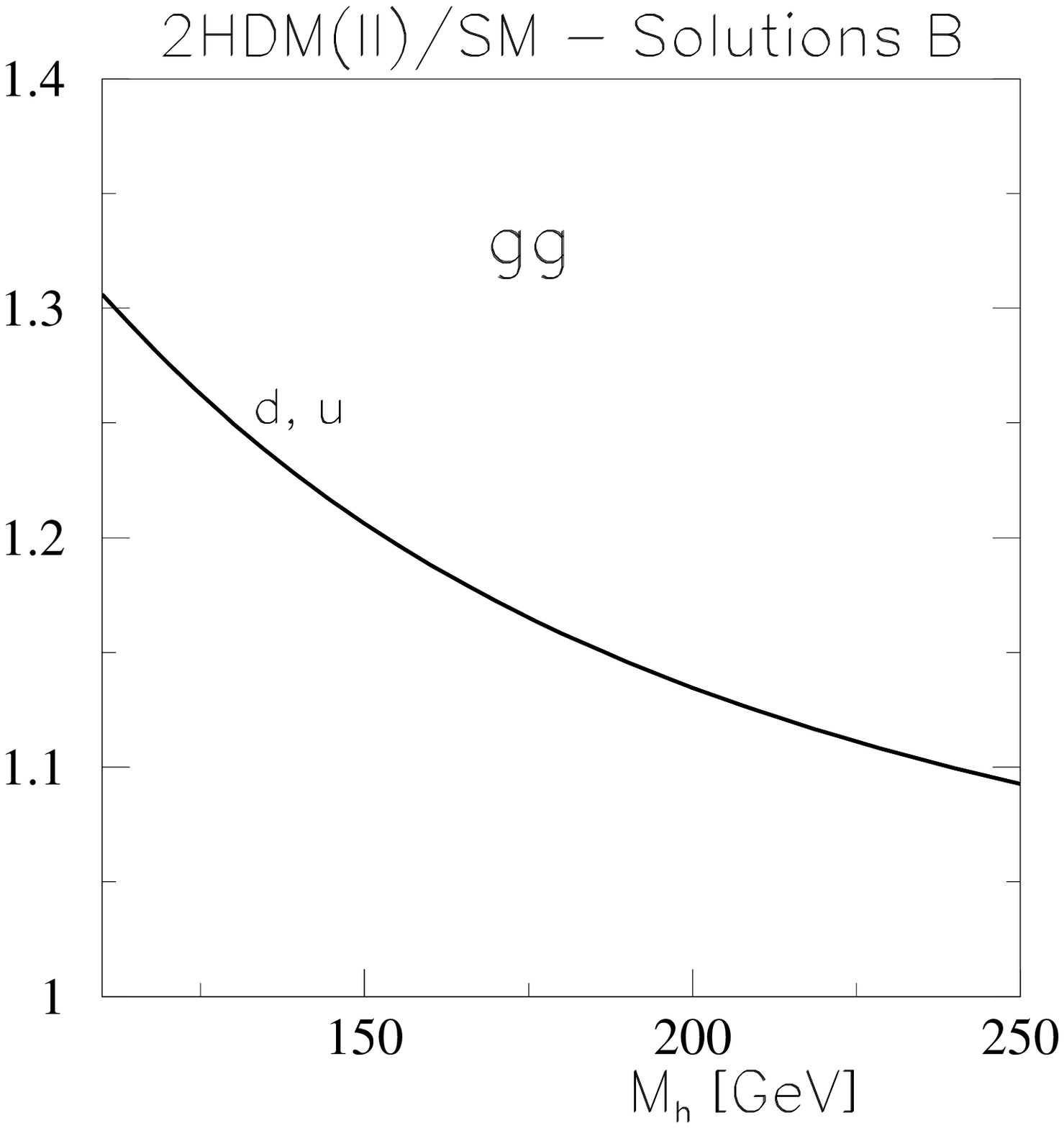}}}
\end{picture}
\vspace*{4mm}
\caption{
Ratios of the Higgs boson decay
widths in the SM-like 2HDM~(II) and the SM as functions of $M_h$.
{\it Left panel}:
$h\to\gamma\gamma$ decay widths, solutions A, for $M_{H^\pm}=800~\text{GeV}$ 
and $\mu/\sqrt{2}=xM_{H^\pm}$.
{\it Right panel}: $h\to gg$, solutions~B.}
\end{center}
\end{figure}

These deviations from unity are large enough that
the form of the 2HDM potential {(large or small $\mu$)}
can be tested at a $\gamma\gamma$ Collider \cite{Aguilar-Saavedra:2001rg}.

Also, the loop-induced couplings to two gluons may differ
from those of the SM-value, but this occurs only for Solutions~B.
This effect is also illustrated in Fig.~1 (right panel).

\noindent
{\large\bf Acknowledgements}\\
{We are grateful  to the CERN, TH Division  for kind hospitality 
and support during work on this paper. 
We thank Maria Jose Herrero Solans for informing us about
ref.~\cite{Ciafaloni:1996ur}.
I.G. is also grateful to the Landau--Centro Volta Network and 
to Prof. O. Panella for
the warm hospitality during a stay at INFN Perugia when the final version
of this paper was prepared.
This work is supported in part by RFBR grants 
02-02-17884 and 00-15-96691 and INTAS 00-00679, by the 
Polish Committee for Scientific Research (grant No. 5 P03B12120 (2002)) and 
by the European Community's
Human Potential Programme under contract HPRN-CT-2000-00149 Physics
at Colliders, and by the Research Council of Norway.}

\end{document}